\documentclass{article}

\usepackage[dblblindworkshop, final]{neurips_2025_libribrain_competition}
\usepackage{algorithm}
\usepackage{algpseudocode}
\usepackage{amsmath}
\usepackage{graphicx}

\usepackage[utf8]{inputenc} 
\usepackage[T1]{fontenc}    
\usepackage{hyperref}       
\usepackage{url}            
\usepackage{booktabs}       
\usepackage{amsfonts}       
\usepackage{nicefrac}       
\usepackage{microtype}      
\usepackage{xcolor}         

\usepackage{graphicx}
\usepackage{wrapfig}
\usepackage{amsmath}
\usepackage{bm}
\usepackage{multirow}
\usepackage[table]{xcolor}
\usepackage[title]{appendix}
\usepackage{float}
\usepackage{natbib}

\newcommand{\std}[1]{\small{$\pm$#1}}

\workshoptitle{LibriBrain Competition 2025}

\title{MEGState: Phoneme Decoding from Magnetoencephalography Signals}

\author{Shuntaro Suzuki\textsuperscript{1\thanks{Equal Contribution.}}, Chia-Chun Dan Hsu\textsuperscript{2*}, Yu Tsao\textsuperscript{2}, Komei Sugiura\textsuperscript{1} \\
\textsuperscript{1}Keio University, Japan \\
\textsuperscript{2}Research Center for Information Technology Innovation, Academia Sinica, Taiwan}

\begin{document}

\maketitle

\begin{abstract}
\vspace{-1em}
Decoding linguistically meaningful representations from non-invasive neural recordings remains a central challenge in neural speech decoding.
Among available neuroimaging modalities, magnetoencephalography (MEG) provides a safe and repeatable means of mapping speech-related cortical dynamics, yet its low signal-to-noise ratio and high temporal dimensionality continue to hinder robust decoding.
In this work, we introduce MEGState, a novel architecture for phoneme decoding from MEG signals that captures fine-grained cortical responses evoked by auditory stimuli.
Extensive experiments on the LibriBrain dataset demonstrate that MEGState consistently surpasses baseline model across multiple evaluation metrics.
These findings highlight the potential of MEG-based phoneme decoding as a scalable pathway toward non-invasive brain–computer interfaces for speech.
\end{abstract}

\section{Introduction}
\vspace{-1em}
Decoding speech representations from brain activity holds considerable promise for restoring communication in individuals with paralysis or severe speech impairments~\citep{moses2021neuroprosthesis}.
Recent advances in invasive brain–computer interfaces~\citep{ECoGDCNet, Cortical-SSM} have enabled continuous speech reconstruction from intracranial recordings, achieving word error rates below 5\% for vocabularies exceeding 100,000 words~\citep{card2024accurate}.
However, their reliance on neurosurgical implantation limits scalability and clinical feasibility.
In contrast, non-invasive approaches such as magnetoencephalography (MEG) offer a safe and repeatable alternative for probing speech-related neural activity.
Nevertheless, decoding linguistically meaningful information from MEG remains challenging due to its low signal-to-noise ratio, high temporal resolution, and sparse neural representations~\citep{yang2024mad, yang2024neuspeech}.

In this work, we introduce MEGState, a novel architecture designed for phoneme classification from MEG signals.
The model integrates two complementary components: (i) a Multi-Resolution Convolution module that captures fine-grained temporal dynamics of phoneme-evoked cortical responses, and (ii) a Sensor-wise SSM that captures long-range temporal dependencies across individual sensors. 
This design effectively mitigates the challenges of MEG’s sparsity and high sampling rate, allowing the model to capture both local and global neural dynamics.
Comprehensive experiments on the LibriBrain dataset demonstrate that MEGState consistently surpasses baseline method across multiple evaluation metrics.
\section{Method}
\vspace{-1em}
\subsection{Preliminaries}
\vspace{-0.5em}
\paragraph{State space models.}
Recent progress in state space models (SSMs)~\citep{S4, Mamba2} shows that they can outperform prevailing architectures, most notably Transformers, on a wide range of sequence modeling tasks.
Grounded in control theory~\citep{KalmanFilter}, SSMs provide a mapping from inputs $\mathbf{x}(t)\in\mathbb{R}^P$ to outputs $\mathbf{y}(t)\in\mathbb{R}^P$ through latent states $\mathbf{h}(t)\in\mathbb{R}^Q$ governed by
\begin{align}
    \label{eq:mimo_ssm}
    \frac{d\mathbf{h}(t)}{dt} = \mathbf{A}\mathbf{h}(t) + \mathbf{B}\mathbf{x}(t) ,\qquad \mathbf{y}(t) = \mathbf{C}\mathbf{h}(t) + \mathbf{D}\mathbf{x}(t),
\end{align}
where $\mathbf{A}\!\in\!\mathbb{R}^{Q\times Q}$ is the state matrix and $\mathbf{B}\!\in\!\mathbb{R}^{Q\times P}$, $\mathbf{C}\!\in\!\mathbb{R}^{P\times Q}$, $\mathbf{D}\!\in\!\mathbb{R}^{P\times P}$ are input/output projections.
Among SSM variants, S5~\citep{S5} has proved especially effective for modeling continuous signals.
S5 sets $\mathbf{A}$ to the HiPPO\mbox{-}N matrix~\cite{S4} to capture long-range temporal dependencies. 
Since HiPPO\mbox{-}N is real symmetric, it admits the diagonalization $\mathbf{A}=\mathbf{V\Lambda V}^{-1}$, which yields the decoupled form
\begin{align}
    \label{eq:mimo_ssm2}
    \frac{d\tilde{\mathbf{h}}(t)}{dt} = \mathbf{\Lambda}\mathbf{\tilde{h}}(t) + \mathbf{\tilde{B}}\mathbf{x}(t),\qquad\mathbf{y}(t)=\mathbf{\tilde{C}}\mathbf{\tilde{h}}(t) + \mathbf{D}\mathbf{x}(t),
\end{align}
where $\tilde{\mathbf{h}}(t)=\mathbf{V}^{-1}\mathbf{h}(t)$, $\tilde{\mathbf{B}}=\mathbf{V}^{-1}\mathbf{B}$, and $\tilde{\mathbf{C}}=\mathbf{CV}$. 
Moreover, introducing a timescale vector $\mathbf{\Delta}\in\mathbb{R}+^{Q}$ and applying zero-order hold discretization~\cite{zero-hold} gives the recurrence
\begin{align}
    \label{eq:mimo_ssm3}
    \mathbf{\tilde{h}}_t = \mathbf{\bar{\Lambda}}\mathbf{\tilde{h}}_{t-1} + \mathbf{\bar{B}}\mathbf{x}_t,\qquad\mathbf{y}_t=\mathbf{\bar{C}}\mathbf{\tilde{h}}_t + \mathbf{\bar{D}}\mathbf{x}_t,
\end{align}
where $\bar{\mathbf{\Lambda}}=\mathrm{exp}(\mathbf{\Lambda}\mathbf{\Delta})$, $\bar{\mathbf{B}}=\mathbf{\Lambda}^{-1}\left(\bar{\mathbf{\Lambda}}-\mathbf{I}\right)\tilde{\mathbf{B}}$, $\bar{\mathbf{C}}=\tilde{\mathbf{C}}$, $\bar{\mathbf{D}}=\mathbf{D}$.
In practice, we take $\mathbf{D}$ to be diagonal and learn $\mathrm{diag}(\mathbf{\Lambda})$, $\tilde{\mathbf{B}}$, $\tilde{\mathbf{C}}$, $\mathrm{diag}(\mathbf{D})$, and $\mathbf{\Delta}$.

\subsection{Model Architecture}
\vspace{-0.5em}
\begin{figure}[t]
    \centering
    \includegraphics[width=\linewidth]{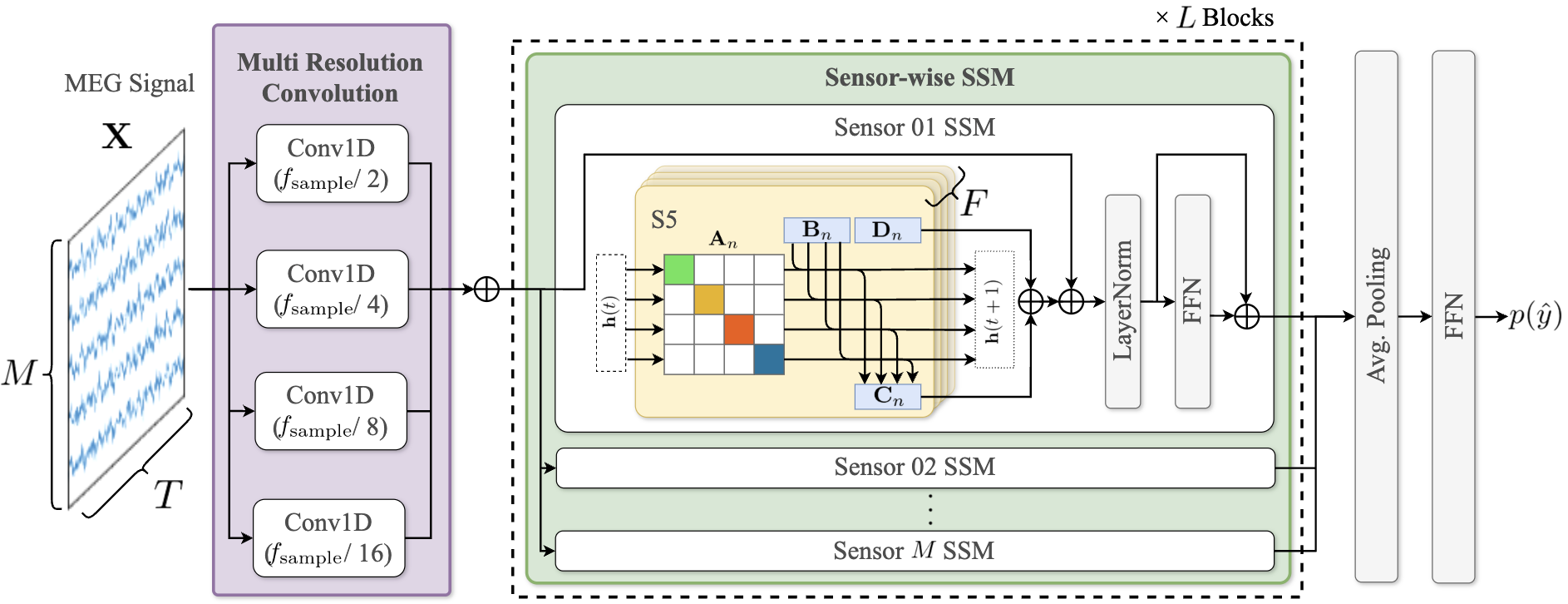}
    \caption{
    Model architecture of the proposed MEGState. Given a MEG signal, Mutli Resolution Convolution module extracts local temporal dependencies while the sensor-wise SSM models global spatial and temporal dependencies.
    }
    \label{fig:model}
    \vspace{-0.5em}
\end{figure}

The overall architecture of the proposed model is depicted in Figure~\ref{fig:model}.
It comprises two primary modules: Multi-Resolution Convolution and Sensor-wise SSM.
Given a MEG sample $\mathbf{X}\in\mathbb{R}^{M\times T}$, where $M$ and $T$ represent the number of sensors and the sequence length, respectively, the network proceeds as follows.

First, the Multi-Resolution Convolution module extract fine-grained local temporal structures reflecting distinct cortical responses evoked by different phonemes.
It consists of four parallel one-dimensional convolutional layers with kernel sizes of $f_\text{sample}/2$, $f_\text{sample}/4$, $f_\text{sample}/8$, and $f_\text{sample}/16$,
where $f_\text{sample}$ denotes the native sampling rate.
The outputs from these convolutional layers are concatenated to yield ${\mathbf{H}}\in\mathbb{R}^{F\times M\times T}$, where $F$ represents the frequency feature dimension.

Next, the Sensor-wise SSM module models global temporal dependencies in a sensor-specific manner.
To handle the sparsity and high temporal resolution of MEG signals, we extend S5~\citep{S5}, a variant of SSM well suited for modeling high-dimensional multivariate time series.
The module comprises $L$ hierarchically organized blocks, and the output $\tilde{\mathbf{H}}^{(l)}\in\mathbb{R}^{F\times M\times T}$ from the $l$-th block $(l=1,\ \ldots,\ L)$ is obtained as follows,
where $\tilde{\mathbf{H}}^{(0)}=\mathbf{H}$:
\begin{align}
    \mathbf{H^\prime}^{(l-1)} = \text{LayerNorm}\left(\text{SSM}\left(\mathbf{\tilde{H}}^{(l-1)}\right) + \mathbf{\tilde{H}}^{(l-1)}\right),\\
    \tilde{\mathbf{H}}^{(l)} = \text{LayerNorm}\left(\text{FFN}\left(\mathbf{H^\prime}^{(l-1)}\right) + \mathbf{H^\prime}^{(l-1)}\right).
\end{align}
Here, $\text{SSM}(\cdot)$, $\text{LayerNorm}(\cdot)$, and $\text{FFN}(\cdot)$ denote the S5 layer, layer normalization, and feed-forward network, respectively.
Subsequently, the output $\tilde{\mathbf{H}}^{(L)}$ from the Sensor-wise SSM is aggregated to produce the predicted phoneme probability $p(\hat{y})$ corresponding to $\mathbf{X}$ as follows:
\begin{align}
p(\hat{y}) = \text{FFN}\left(\text{AvgPool}\left(\tilde{\mathbf{H}}^{(L)}\right)\right),
\end{align}
where $\text{AvgPool}(\cdot)$ denotes average pooling over the temporal dimension, following the standard SSM architectures~\citep{S4}.
The model is trained using the cross-entropy loss function.
\section{Experiments}
\vspace{-0.5em}
\subsection{Dataset and Pre-procesing}
\vspace{-0.5em}
In this experiment, we used the publicly available LibriBrain dataset~\cite{LibriBrain}, which contains MEG recordings of a single participant listening to audiobook narrations of Sherlock Holmes.
The recordings were acquired with a MEGIN Triux\textsuperscript{TM} Neo system using 306 sensors at 1 kHz, yielding 52.32 hours of data.
Preprocessing involved Maxwell filtering~\citep{Maxwell} and Signal Source Separation to remove sensor noise and external magnetic interference.
A notch filter (50 Hz), Butterworth band-pass filter (0.1–125 Hz), and downsampling to 250 Hz were further applied.
The dataset provides 39 ARPAbet-based phoneme labels temporally aligned with the auditory stimuli, where each label corresponds to a 0.5 s MEG segment.
It was divided into training, validation, and test sets comprising 1,488,392, 11,289, and 12,051 samples, respectively.
Models were trained, tuned, and evaluated on these splits, and further tested on the LibriBrain leaderboard set provided in the 2025 PNPL Competition~\citep{PNPL}.

\subsection{Implementation Details}
\vspace{-0.5em}
\begin{algorithm}[t]
\caption{Sampling training data}
\begin{algorithmic}
\State \textbf{Input:} Dataset $(\mathcal{X}, \mathcal{Y}) = \{(\boldsymbol{X}_i, y_i)\}_{i=1}^{N}$
\State \textbf{Output:} Training sample $(\tilde{\boldsymbol{X}}, \tilde{y})$

\State Randomly sample two labels $y_1 \sim \mathrm{Uniform}(\mathcal{Y})$, $y_2 \sim \mathrm{Uniform}(\mathcal{Y})$
\State $\mathcal{I}_1 \gets \{\,i \in \{1, \dots, N\} \mid y_i = y_1\,\}$ {\textcolor{blue}{\Comment{All indices with label $y_1$}}}
\State $\mathcal{I}_2 \gets \{\,i \in \{1, \dots, N\} \mid y_i = y_2\,\}$ {\textcolor{blue}{\Comment{All indices with label $y_2$}}}
\State Randomly sample subsets $\mathcal{I}_1' \subset \mathcal{I}_1$, $\mathcal{I}_2' \subset \mathcal{I}_2$ s.t. $|\mathcal{I}_1'| = |\mathcal{I}_2'| = N'$
\State $\bar{\boldsymbol{X}}_1 \gets \text{AVERAGE}_{i \in \mathcal{I}_1'}(\boldsymbol{X}_i)$
\State $\bar{\boldsymbol{X}}_2 \gets \text{AVERAGE}_{i \in \mathcal{I}_2'} (\boldsymbol{X}_i)$
\State $\tilde{\boldsymbol{X}} \gets \alpha \bar{\boldsymbol{X}}_1 + (1-\alpha)\bar{\boldsymbol{X}}_2$ {\textcolor{blue}{\Comment{Mixup augmentation}}}
\State$\tilde{y} \gets \alpha y_1 + (1-\alpha)y_2$
\end{algorithmic}
\label{algo:sampling}
\end{algorithm}


During training, we leverage smoothing and data augmentation in the sampling of the training samples to improve the signal-to-noise ratio (SNR) of MEG signals and mitigate phoneme-label imbalance.
Concretely, at each sampling step within a training epoch, we independently draw two phoneme labels $y_1$ and $y_2$ uniformly from the MEG dataset $(\mathcal{X},\mathcal{Y})=\{(\bm{X}_i,y_i)\}^N_{i=1}$. 
For each label $y_k\ (k\in \{1,2\})$, we construct the index set of all samples with that label, $\mathcal{I}_k=\{i\in\{1,\ \ldots,\ N\}\mid y_i=y_k\}$, and uniformly sample a subset $\mathcal{I}_k'\subset \mathcal{I}_k$ of size $N'$. 
We then form a class-conditional prototype by averaging the corresponding MEG inputs, $\bar{\bm{X}}_k=\frac{1}{N^\prime}\Sigma_{i\in\mathcal{I}_k}\bm{X}_i$, which acts as a denoised representative and thus enhances the SNR. 
Furthermore, by introducing a mixing coefficient $\alpha\in[0,1]$, we address label imbalance (see Sec.~\ref{sec:results}) via a mixup-style convex combination~\citep{mixup} of both the prototypes and their labels:
\begin{align}
    \tilde{\bm{X}} = \alpha\bar{\bm{X}}_1 + (1-\alpha)\bar{\bm{X}}_2,\quad 
    \tilde{y} = \alpha\bar{y}_1 + (1-\alpha)\bar{y}_2.
\end{align}
The overall sampling procedure is summarized in Algorithm~\ref{algo:sampling}.

We trained the model with the AdamW optimizer ($\beta_1=0.9,\ \beta_2=0.999$) and a learning rate of $1.0\times10^{-4}$. 
The batch size was 32, and training proceeded for 50 epochs. 
For the Sensor-wise SSM module, we set the block size to $L=2$.
For sampling training data, we set the prototype size of $N^\prime=100$ and a mixing coefficient of $\alpha=0.5$. 
\begin{table}[t]
    \centering
    \caption{Performance comparison on the LibriBrain~\citep{LibriBrain} test set and leaderboard set. \textbf{Bold} values denote the best performances, while $\dagger$ indicates statistical significance compared to the baseline method ($p < 0.05$). Multi-Resol Conv. indicates Mutli-Rresolution Convolution module.}
    \label{tab:ablation-module}
    \renewcommand{\arraystretch}{1}
    \resizebox{\textwidth}{!}{
    \begin{tabular}{l c c c c}
      \toprule
      \multirow{2}{*}{\textbf{Model}}& \multicolumn{3}{c}{\textbf{Test Set}}& \textbf{Leaderboard Set}\\
      \cmidrule(l{1mm}r{1mm}){2-4}
      \cmidrule(l{1mm}r{1mm}){5-5}
       & \bf{Acc.~[\%]~$\uparrow$} & \bf{Kappa~[\%]~$\uparrow$} & \bf{Macro-F1~[\%]~$\uparrow$} & \bf{Macro-F1~[\%]~$\uparrow$}\\
      \midrule
      Baseline~\citep{LibriBrain} &38.80\std2.40 &45.71\std0.74 &34.82\std1.92 &---\\
      Ours (w/o Mulit-Resol. Conv.)& 40.25\std3.15 &37.90\std2.83$^\dagger$ & 34.77\std1.83& ---\\
      Ours (w/o Sensor-wise SSM)& 40.37\std3.20& 49.60\std6.25& 37.18\std4.44& ---\\
      Ours (MEGState)& \textbf{45.53\std1.88}$^\dagger$& \textbf{54.19\std2.42}$^\dagger$& \textbf{41.11\std2.20}$^\dagger$& \textbf{55.74}\ (68.41)\\
      \bottomrule
    \end{tabular}
    \label{table:main_quant}
    }
    \vspace{-1em}
\end{table}

\begin{figure}[t]
    \centering
    \includegraphics[width=\linewidth]{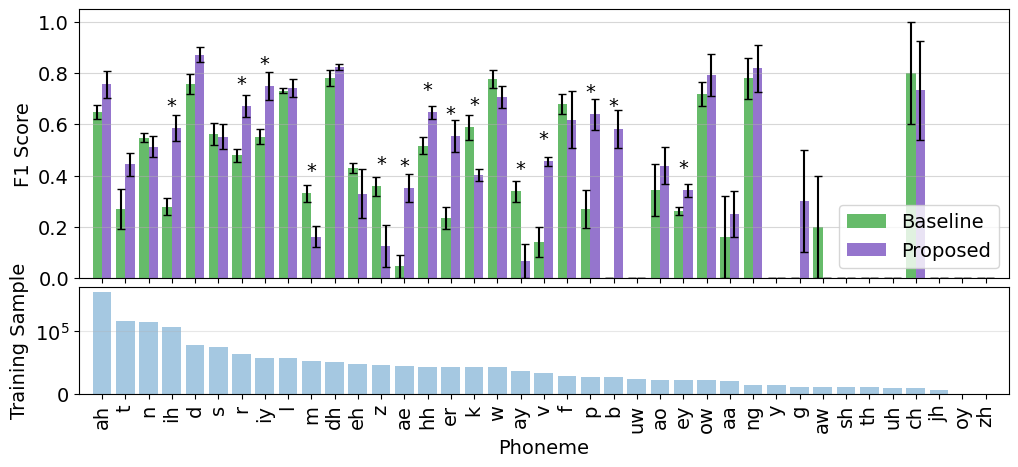}
    \vspace{-2em}
    \caption{
    Quantitative comparison across phonemes. The upper panel shows the macro-F1 for each phoneme, while the lower panel indicates the number of training samples per phoneme. Error bars represent the standard error of the mean, and $*$ denotes statistical significance ($p < 0.05$).
    }
    \label{fig:phoneme_f1}
    \vspace{-1em}
\end{figure}

\section{Results}
\vspace{-0.5em}
Table~\ref{table:main_quant} presents the quantitative comparison between the proposed and baseline methods.
Values in the table represent the mean and standard deviation obtained from over five distinct random seeds.

On the test set, the proposed method achieved balanced accuracy, Cohen’s Kappa, and macro-F1 of 45.53\%, 54.19\%, and 41.11\%, respectively, surpassing the baseline method by 6.73, 8.48, and 6.29 points.
Here, all improvements were statistically significant ($p < 0.05$).
Moreover, ablation studies revealed that removing either the Multi-Resolution Convolution module or the Sensor-wise SSM consistently degrades performance across all metrics on the test set, indicating that both modules contribute substantially to phoneme classification from MEG signals.

On the leaderboard set, the proposed method achieved a macro-F1 of 55.74\%.
Notably, the final leaderboard submission employed an ensemble strategy that selected the most probable label from predictions of five independently trained models, resulting in a higher macro-F1 of 68.41\%.

Figure~\ref{fig:phoneme_f1} further presents the phoneme-wise quantitative comparison.
Each bar shows the macro-F1 score for individual phonemes, ordered by the number of training samples.
As shown, the proposed method outperformed the baseline in terms of macro-F1 on 19 phonemes and achieved statistically significant improvements on 10 phonemes ($p < 0.05$).

\label{sec:results}

\section{Conclusion}
\vspace{-0.5em}
In this work, we introduced MEGState, a novel architecture for phoneme decoding from MEG signals.
By integrating a Multi-Resolution Convolution module to capture fine-grained local temporal dynamics and a Sensor-wise SSM to model long-range dependencies across individual sensors, our approach effectively mitigates the challenges of MEG signal sparsity and high temporal resolution.
Comprehensive experiments on the LibriBrain dataset demonstrated that MEGState consistently outperformed the baseline across multiple evaluation metrics.

\newpage

\bibliographystyle{plainnat}
\bibliography{refs}

\end{document}